**PAPER • OPEN ACCESS**

# Nonlinear polarization effect of functionalized graphene quantum dots

To cite this article: Setianto and I Made Joni 2022 J. Phys.: Conf. Ser. **2344** 012006

View the article online for updates and enhancements.

## You may also like

- Global quantum discord in the Lipkin–Meshkov–Glick model at zero and finite temperatures
  Jia Bao, Yan-Hong Liu and Bin Guo

- Optical Properties Alteration and Photo-Voltaic Applications of Nitrogen-Doped Graphene Quantum Dots
  Md Tanvir Hasan, Roberto Gonzalez-Rodriguez and Anton V Naumov

- Protecting geometric quantum discord via partially collapsing measurements of two qubits in multiple bosonic reservoirs
  Xue-Yun Bai, , Su-Ying Zhang et al.

# Nonlinear polarization effect of functionalized graphene quantum dots

**Setianto**[1,2,3 a] **and I Made Joni**[2,3,b*]

[1]School of Biotechnology, Universitas Padjadjaran, Jl. Raya Bandung-Sumedang KM. 21 Jatinangor, Sumedang, Indonesia, 45363

[2]Functional Nano Powder University Center of Excellence, Padjadjaran University, Jl. Raya Bandung-Sumedang KM 21, Jatinangor, 45363, Jawa Barat, Indonesia

[3]Department of Physics, Universitas Padjadjaran, Jl. Raya Bandung-Sumedang km. 21 Jatinangor, 45363, Sumedang, Indonesia

[a]setianto@phys.unpad.ac.id,　[*b]imadejoni@phys.unpad.ac.id

**Abstract**. Graphene quantum dots (GQDs) are nanoscale structures of graphene with quantum properties and edge effects that give photoluminescence properties. The effect of quantum confinement and differences in the nature of GQD structure makes its optical characteristics highly dependent on the size of the structure. This study explains a few exploratory semi-empirical calculations of nonlinear polarization properties of functionalized GQD (fGQD) three-dimensionally. Based on this, the calculation of the linear polarization and first hyperpolarization was performed by the finite field method, which is based on the expansion of the energy and dipole moment. As a result, the fGQD molecule dominantly has high optical nonlinear properties as indicated by the high $\beta$ values (71 to 4488 a.u.). In general, the first hyperpolarizabilities have a linear relationship with the dipole moments. It was potentially used for the second harmonic imaging microscopy (SHIM) application.

## 1. Introduction

Graphene quantum dots (GQD) represent one to several graphene layers with a size of less than 10 nm. GQD is classified as a relatively new molecule that attracts extraordinary interest because of its have unique in optical [1–6], electrical[7–9], chemical [10–15], and structural properties [16–18]. The optical nonlinearity of organic materials has been investigated and become a subject of extensive research to understand their intrinsic properties and their possible usage for photonic applications [8,19,20] and their potential use in photonic applications [21]. In this context, it is imperative to use the nonlinear optical properties of GQD materials for applications in optoelectronic components, photonics, and microscopy imaging [22–25].

All molecular crystals with a non-centrosymmetric geometry are potential candidates for nonlinear optical materials (NLO) because they can show the effect of second harmonic generation (SHG). The Second Harmonic Generation (SHG) effect was demonstrated for the first time by Peter Franken et al.

in 1961 [26]. A year later, in 1962, researchers at Harvard explained the initial formulation of the SHG [27]. The comprehensive rules for understanding the interaction of light in nonlinear media are described using Maxwell's equations at the planar interface between linear and nonlinear media. In materials with a non-centrosymmetric structure, the second harmonic coefficient is not zero, resulting in nonlinear polarization and generated SHG light. It has been observed that some non-centrosymmetric organic dyes can produce second harmonic generation [28]. However, most organic dyes also produce collateral fluorescence along with a signal to produce the second harmonic. So far, only two classes of organic dyes do not produce collateral fluorescence and only work in the second harmonic generation [9]. The characterization of nanostructured materials can also be carried out with the second harmonic microscope [29].

Computationally, hexagonal graphene quantum dot (GQD) which is induced by an external electric field will generate a nonlinear optical effect (NLO). As a result, The different anisotropy properties are shown when GQD is induced by an external electric field in different direction fields so that the first hyperpolarization value depends on the direction of the electric field [7]. Theoretically, it shows that the NLO response of the hexagonal GQD molecule can be electrically tuned.

In this study, we explain a few exploratory semi-empirical calculations of nonlinear polarization properties of functionalized GQD (fGQD) three-dimensionally. Based on this, the calculation of the linear polarization and first hyperpolarization was performed by the finite field method, which is based on the expansion of the energy and dipole moment. The calculation of the values and components of the dipole moment ($\mu$), linear polarization ($\alpha$), and first hyperpolarization ($\beta$) in this study was performed using the AM1 semi-empirical method [30] in the MOPAC program [31]. The SHG effect on GQD will be visualized using a two-dimensional mesh plot to simulate the SHG light as a function of the external electric field-induced medium and polarization. We also analyze the SHG effect on polar coordinates as an illustration of the pattern of changes in the distribution of nonlinear polarization.

## 2. Material and Method

*2.1. Material*

In the GQD state, all four rings of pyrene structure are changed into nine membered rings (Figure 1a) and denoted as pristine (pGQD). The system is called fGQD when pGQD is functionalized by adding five hydroxyl groups (-OH) and a methyl group (-$CH_3$) at the edges and 4 oxygen atoms in the middle of the surface (Figure 1b).

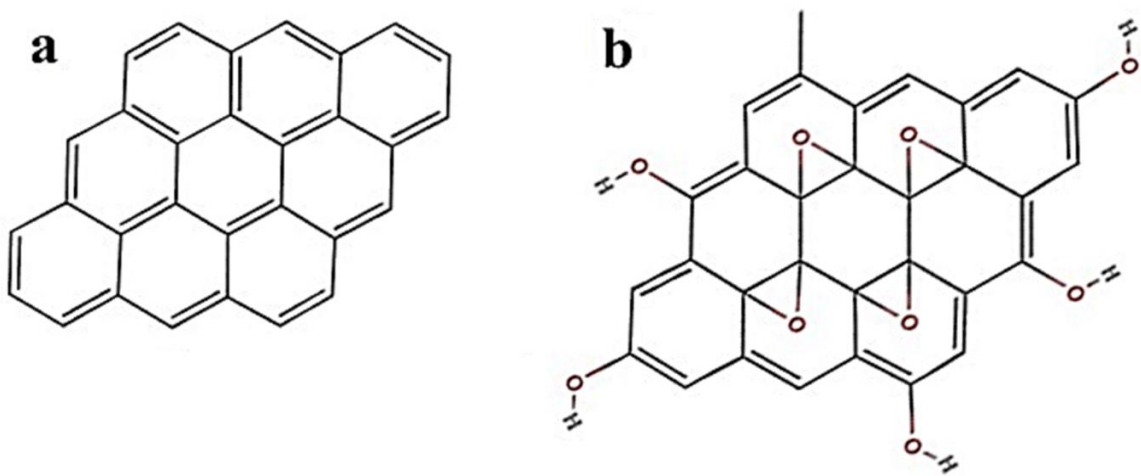

**Figure 1.** The pristine and functionalized Graphene Quantum Dot structure: a) pGQD and b) fGQD.

*2.2. Computational method*

The GQD molecular geometry was optimized using the AM1 semi-empirical method [30]. For all structures shown in Figure. 1, the planar configuration is assumed as the starting point for the calculation of the geometric optimization. Since our ultimate goal is to compute the first hyperpolarizability value of fGQD, we did not consider the explicit contribution of the solvent effect in the first analysis of this minimal energy structure. In order to study the nonlinear polarization effect, the group positions of the donor and acceptor molecules at the para position are selected [32]. As the donor group, we selected the dimethylamine (N(CH$_3$)$_2$) molecule and we used the hydrogen (H) atom, the hydroxyl group (OH), and the nitrogen dioxide (NO$_2$) molecule as acceptor groups (Figure. 2). We performed semi-empirical calculations with MOPAC standard parameters [33–35] with the exception of the convergence criterion, which uses a maximum step size of 0.05.

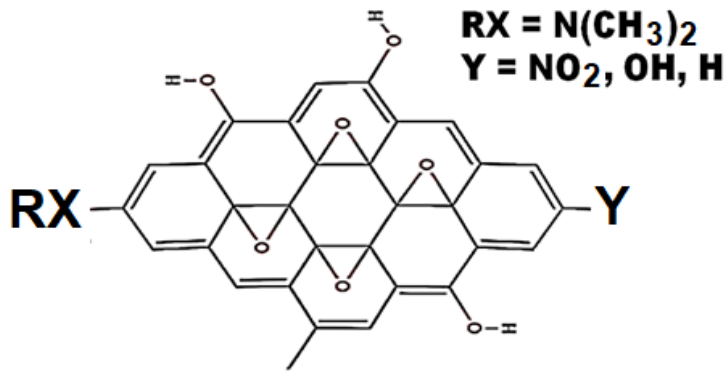

**Figure 2.** The structure of fGQD for nonlinear polarization calculations, with dimethylamine molecule as donor group and a hydrogen atom (fGQD-shg0), the hydroxyl group (fGQD-shg1), and the nitrogen dioxide molecule (fGQD-shg2) as acceptor groups, respectively.

In the nonlinear polarization properties, the molecular polarization is approximated as a dipole due to an electric field:

$$\mu_i(E) = \mu_i^0 + \sum_j \alpha_{ij} E_j + \sum_{jk} \beta_{ijk} E_j E_k + \cdots \quad (1)$$

Based on this, the calculation of the linear polarization and first hyperpolarizability was performed by the finite field method, which is based on the expansion of the energy and dipole moment [35]. The complete equation for calculating the magnitude of the first hyperpolarizability is given as:

$$\beta_{total} = \left[ \begin{array}{c} (\beta_{xxx} + \beta_{xyy} + \beta_{xzz})^2 + \\ (\beta_{yyy} + \beta_{yzz} + \beta_{yxx})^2 + \\ (\beta_{zzz} + \beta_{zxx} + \beta_{zyy})^2 \end{array} \right]^{1/2} \quad (2)$$

To illustrate the nonlinear effect of fGQD, where $\chi^{(2)}$ is a second-order susceptibility, we use an external electric field that is induced at the medium $E = E_o \cos(\omega t + kr)$. The nonlinear polarization phenomenon can be seen using the equation:

$$P = \chi^{(1)} E_0 \cos(\omega t + kr) + \frac{1}{2}\chi^{(2)} E_0^2 [1 + \cos(2\omega t + 2kr)] \quad (3)$$

This equation shows the frequency component caused by nonlinear polarization. The second term indicates a frequency of 2ω, referred to as the second harmonic generation (SHG) which only appears in non-centrosymmetric media [36–40]. In general, the interaction of light with the medium can be viewed as a dielectric to an electric field. This approach is often represented by an approximation of the dipole moment, $\mu = -e\,r$, where $e = 1.6 \times 10^{-19}$ *Coulomb* and *r* is the distance of the induced field. This leads to a polarization which is the result of dipole induction,

$$P = N\,\mu \quad (4)$$

with *N* is electron density in the medium. Based on equation (1), (3), and equation (4) then:

$$\chi^{(1)}_{ij} = N\alpha_{ij} \tag{5}$$

$$\chi^{(2)}_{ijk} = N\beta_{ijk} \tag{6}$$

show the relationship between susceptibility and polarizability.

## 3. Result and Discussion

### 3.1. First Hyperpolarizability of GQD

The AM1 calculation results for the corresponding dipole moment (μ), linear polarization (α), and first hyperpolarization (β) in the static field for the pristine GQD (pGQD) and functionalized GQD (fGQD) is given in Table 1. We find that the linear polarizability (α) values only change slightly between 380 and 444 a.u. and the first hyperpolarizability (β) values change significantly from 0 to 4488 a.u. It is essential to note that the pGQD molecule does not have a nonlinear polarization effect with small β values (close to zero). As for the fGQD molecule, it dominantly has high nonlinear polarization as indicated by the high β values (71 to 4488 a.u.). In general, the first hyperpolarizabilities have a linear relationship with the dipole moments.

**Table 1.** The AM1 calculated total first hyperpolarizability for pristine and three functionalized GQD

| System | μ (debye) | α (a.u.) | β (a.u.) |
|---|---|---|---|
| pGQD | 0 | 380 | -0.067 |
| fGQD-shg0 | 2.53 | 393 | 71 |
| fGQD-shg1 | 3.43 | 423 | 699 |
| fGQD-shg2 | 9.37 | 444 | 4488 |

### 3.2. Second Harmonic Generation (SHG) of GQD

In the medium that is induced by an electric field generated by the laser, the polarization response of the material is determined by an equation (3). The nature of the light wave can be explained by the oscillation of the electromagnetic field with the spread of the field strength $E(r, t)$ in varying space and time [41,42]. Figure 3 is a visualization of the effect of nonlinear polarization on GQD molecules due to the electric field. For pristine (pGQD), there is no SHG effect because the dipole moment value is zero (Figure 3a). So that it produces a circular curve at polar coordinates. This means that pGQD produces linear polarized light. For the functionalized GQD (fGQD-shg0), there is a relatively small SHG effect which is indicated by a lower beta value than the alpha value (Figure 3b). So that it produces two circular curves in one sphere at the polar coordinates. It can be interpreted that there are two linearly polarized lights. In Figure 3c-d, the effect of SHG is increasingly dominant in functionalized GQD molecules (fGQD-shg1 and fGQD-shq2). This can be seen with the appearance of circular polarization separated by a certain distance. The β value is also higher than the α value. So that the effect of nonlinear polarization can be clearly seen by the presence of two elliptical curves that are opposite to each other at the polar coordinates. This curve is often referred to as lemniscate. The interpretation of the curve is that light is non-linearly polarized. In other words, the effect of SHG dominance occurs on the fGQD-shg2 molecule.

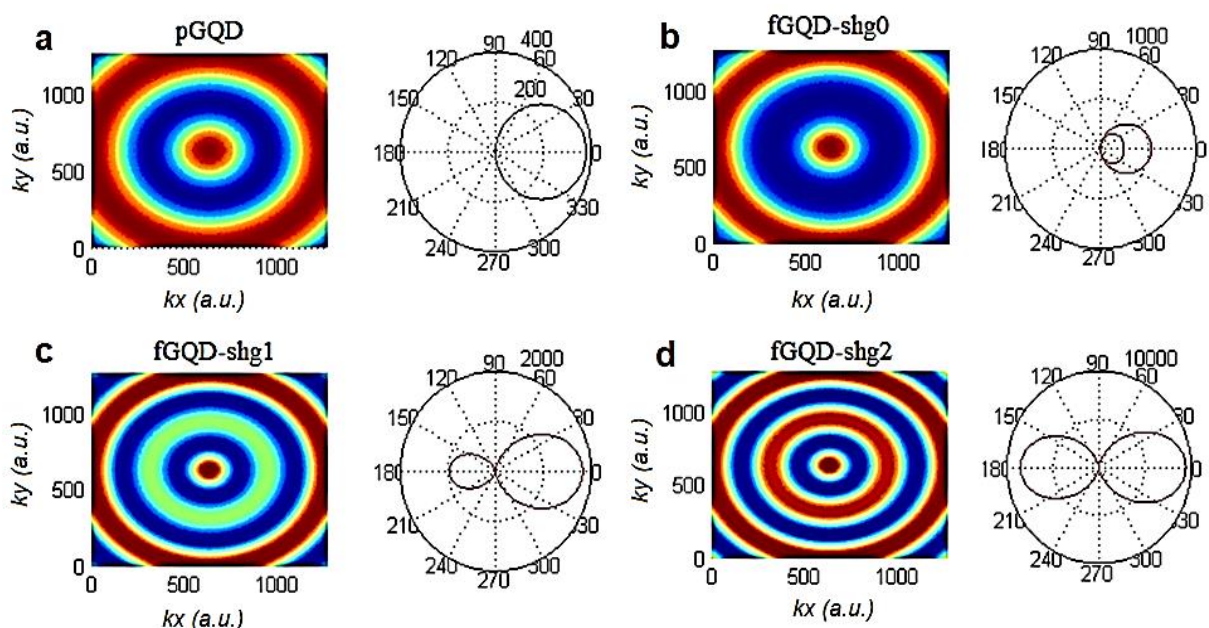

**Figure 3.** Visualization of the effect of nonlinear polarization on GQD molecules due to the external electric field-induced; a) pGQD b) fGQD-shg0 c) fGQD-shg1 and d) fGQD-shg2

## 4. Conclusion

We have presented semi-empirical calculations for investigating the first hyperpolarizability value of fGQD with different functional groups. The fGQD molecular structure induced by an electric field leads to a change in the value of the dipole moment associated with the effect of nonlinear polarization, and this is indicated by the circularly polarized light. The SHG effect of fGQD-shg2 is very clear and potentially used for the second harmonic imaging microscopy (SHIM) application.

**Acknowledgement**

This work is supported by Academic Leadership Grant and RDDU Project 2020 with contract number: 1427/UN6.3.1/LT/2020